\documentclass{article}
\usepackage[top=0.85in,left=0.75in,footskip=0.75in,marginparwidth=2in]{geometry}
\usepackage{authblk}
\usepackage{etoolbox}
\usepackage{booktabs}
\usepackage{multirow} 
\usepackage{float} 
\usepackage{subfigure}
\usepackage{graphicx}
\usepackage{dcolumn}
\usepackage{amsmath}
\usepackage[utf8]{inputenc}

\title{A reconfigurable calibration-free  digital-to-time converter based on a high-speed transceiver}
\author[1]{Dexuan Kong}
\author[1]{Zaiming Fu}
\author[1]{Yujie Deng}
\author[1]{Ruiqi Wang}

\affil[1]{University of Electronic Science and Technology of China, School of Automation Engineering, Chengdu 611731, China}

\begin{document}
\maketitle 

\begin{abstract}
This paper proposes a high-speed transceiver-based method for implementing a digital-to-time converter (DTC). A real-time decoding technique is introduced to inject time information into high-speed pattern data. The stability of the high-speed clock ensures the high precision of the synthesized timing signal without the need for calibration. The reconfigurability of the clock resources provides the DTC with variable resolution and enhanced flexibility for various applications. Based on this approach, a multifunctional DTC is designed to offer both timing sequence and random timing signal functionalities, catering to a wide range of application scenarios. The timing sequence function generates a continuously variable timing signal stream, while the random timing signal function produces random signals with uniformly distributed time intervals. Experimental results, using a Xilinx Kintex-7 FPGA, validate the effectiveness of the proposed methodology. The system achieves a resolution of 100 ps, a dynamic range from 1 ns to 40 $\mu$s, a DNL of -0.02/0.02 LSB, an INL of -0.04/0.03 LSB across the entire range. This approach can be readily adapted to various high-precision timing signal applications.
\end{abstract}

\newcommand\keywords[1]{\textbf{Keywords}: #1}

\section{Introduction}

Time interval signals, also known as timing signals, encode time information in the width of pulse signals. These signals are essential in time-domain measurement systems, including particle physics experiments and time-of-flight (ToF) measurements. Digital-to-time converters (DTCs) have been integral to automatic test equipment (ATE) for many years, finding use in applications such as very large-scale integration (VLSI) functional testing and time-to-digital converter (TDC) evaluation. Recently, DTCs have been increasingly adopted in emerging fields. For instance, in intelligent driving, DTC-based infrared radar detects obstacles by reconstructing reflected signals and measuring ToF, providing precise distance measurements between vehicles and obstacles\cite{radar1,radar2,radar3}. In biomedicine, DTC-based ADCs are employed in electrical stimulation devices like epilepsy suppressors and cardiac pacemakers\cite{med1,med2}. In communications, timing signal measurements are critical for next-generation technologies\cite{5g1,5g2,5g3,5g4,5g5,5g6,5g7,5g8}. Fractional-N phase-locked loops use DTCs to suppress quantization noise and spurious signals. Meanwhile, DTCs play a pivotal role in generating precise timing sequences in quantum computing\cite{quantum1,quantum2,quantum3}.

DTC implementation methods primarily fall into two categories: application-specific integrated circuits (ASICs) and FPGA-based designs. DTC schemes based on ASICs are known for their exceptionally high timing resolution, sometimes reaching sub-picosecond levels. Their size, power consumption, linearity, and overall performance can be optimized through specialized circuit design\cite{asic1,asic2,asic3,asic4}. However, their fixed hardware structure limits their adaptability to different testing requirements. As general-purpose integrated circuits with programmable features, FPGAs provide greater flexibility. FPGA-based DTC designs have gained traction, with numerous innovative architectures proposed in recent years. Common FPGA-DTC implementation approaches include direct counting, Vernier, and delay line methods. The direct counting method generates timing signals by counting system clock cycles, with resolution determined by the clock period\cite{count1,count2,count3}. While simple in structure, its resolution is limited by the operating frequency of FPGA chips. The Vernier method achieves high resolution by measuring differences between two clock signals with closely spaced periods but usually offers a narrow dynamic range\cite{vernier1,vernier2,vernier3}. The delay line method provides high resolution by processing signals through staged delays and multiplexers\cite{delay1,delay2,delay3}. However, it suffers from sensitivity to temperature and voltage variations, requiring complex calibration. 

This paper proposes a novel DTC design based on FPGA high-speed transceivers. The proposed design can achieve highly precise timing signal generation by leveraging the stability of high-frequency clocks from high-speed transceivers. We introduce a real-time encoding algorithm that efficiently converts time interval information into high-speed code patterns to address FPGAs' memory limitations. Building on this, we design a multifunctional DTC capable of outputting both timing signal sequences and random timing signals. The main contributions of this work are as follows:

(1)We propose a high-speed transceiver-based DTC implementation. Parallel data is serialized using real-time encoding to generate precise timing signals without calibration.

(2)We design a multifunctional DTC capable of producing both timing signal sequences and random timing signals.

(3)We implement and validate a 10 Gbps transceiver-based DTC on a Xilinx Kintex-7 FPGA, demonstrating its effectiveness and precision.

\section{Architecture and principle}
\begin{figure}[htp]
\centering
\includegraphics[width=8.5cm]{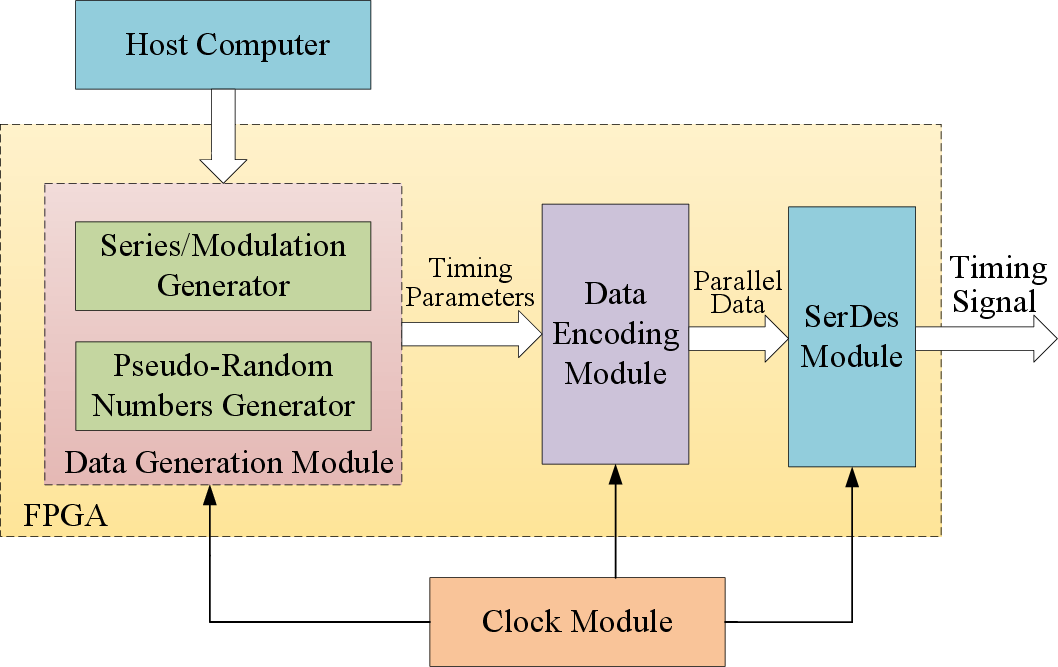}
\caption{Overall structure of the proposed DTC.}
\label{fig:1}
\end{figure}

Fig.\ref{fig:1} shows the structure of the proposed DTC, mainly consisting of FPGA and the clock module. In the figure, the FPGA has three modules: the data generation module, the data encoding module, and the serializer/deserializer(SerDes) module. DTC consists of the data encoding module and SerDes. Depending on the type of signal and parameters needed, the user controls the data generation module through the control computer to generate the timing parameters and input them into the data encoding module. The timing parameter contains information about the type of output signal and parameters. Taking the received timing parameters and the reference clock provided by the clock module as its foundation, the data encoding module generates parallel data and inputs it into the SerDes. Finally, the SerDes converts the parallel data to the desired timing signal. The specific generation process of the timing signal will be described in detail below in the order of the signal flow.

\subsection{Data generation}

This DTC generates single timing signals with interval ranging from 2 ns to 40 $\mu$s. The user specifies the desired pulse width via a control computer, which sends the interval information to the data generation module. The module then calculates and outputs the corresponding timing parameters.
\begin{figure}[htp]
    \centering
    \includegraphics[width=8.5cm]{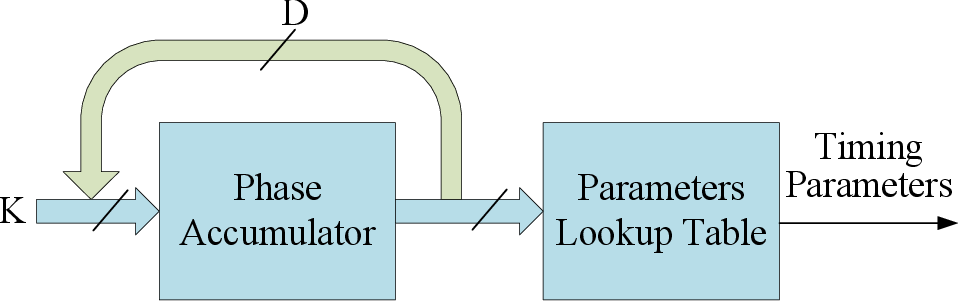}
    \caption{Structure of Sequences Generator}
    \label{fig:2}
\end{figure}

By serializing these single timing signals, a sequence with fixed timing intervals can be created. This DTC supports sequences containing 2 to 1000 fixed intervals. Additionally, automated test equipment often requires timing signals with varying intervals, known as timing sequences. To generate these, the DTC employs a sequence generator module, which determines the timing parameters based on specified interval variations.

The total duration of the high and low levels of a timing signal is referred to as its length. In the FPGA implementation, this can be achieved using a phase accumulator combined with a parameter lookup table, allowing precise control over signal lengths. The structure, illustrated in Fig.\ref{fig:2}, also supports the generation of timing sequences with varying intervals through a timing modulation function.

The user defines parameters such as timing interval variation rules in the lookup table via the control computer. The phase accumulator in this DTC has a resolution of $D$ bits, and its incremental step, known as the frequency control word, is denoted as 
$K$. Driven by a sampling clock, the phase accumulator updates by $K$ at each clock cycle, retrieves data from the parameter lookup table, and outputs the corresponding timing parameters.

\begin{figure}[htp]
    \centering
    \includegraphics[width=8.5cm]{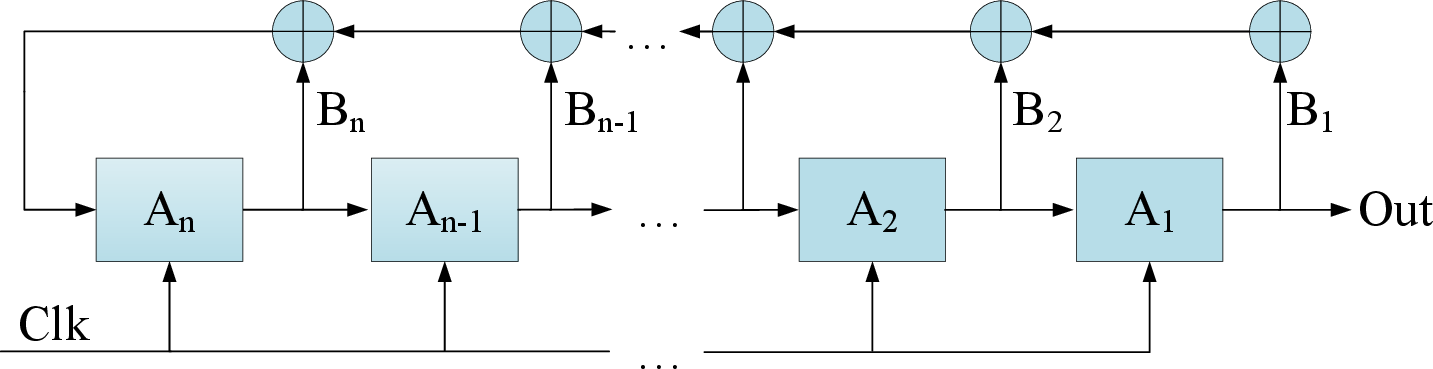}
    \caption{Structure of an n-level LFSR.}
    \label{fig:3}
\end{figure}

The relationship between $L$ (the length of the individual signals in the output timing sequences) and $f_s$ (the clock frequency) is expressed by Eq.(\ref{eq:1}).
\begin{equation}
L  =  \frac{2^{D}}{K \cdot f_{s}}.
\label{eq:1}
\end{equation}

Varying the frequency control word $K$ can change the length of the individual signals in the timing sequences of the output signal. Timing signal with random timing intervals are critical in controlling switching power supplies and motor drives. This DTC device can generate timing signal with random intervals by generating pseudo-random numbers. Due to their simplicity and ease of implementation, linear feedback shift registers(LFSRs) have become the dominant structure for generating pseudo-random numbers in FPGAs. Fig.~\ref{fig:3} shows the structure of an n-level LFSR. A feedback loop is added to an n-stage shift register, which can conduct Exclusive-OR operation on some bits in the shift register and fill the first register with the result. 
\begin{equation}
\begin{aligned}
f(x) &= 1 + B_{n} x^{n} + B_{n-1} x^{n-1} + \cdots + B_{2} x^{2} + B_{1} x \\
     &= \sum_{i=0}^{n} B_{i} x^{i}.
\label{eq:2}
\end{aligned}
\end{equation}

Eq.(\ref{eq:2}) represents the characteristic polynomial of the LFSR. Exclusive-OR operation is denoted as $\oplus$. The $A_{i}$ is the data stored in the $i-th$ register.
The $B_{i}$ is the feedback coefficient. $A_{n+1}$ (the feedback value received at the leftmost side of the LFSR) is shown in Eq.(\ref{eq:3}).
\begin{equation}
A_{n+1} = A_{1}B_{1} \oplus A_{2}B_{2}  \oplus \cdots \oplus A_{n}B_{n}.
\label{eq:3}
\end{equation}

The data output by the LFSR is periodic. If the period of the n-level LFSR is $2^{n-1}$, the generated sequence is called an $m$ sequence. However, the period of an $m$ sequence is limited. Multiple $m$ sequences are used to simultaneously output pseudo-random numbers, which greatly extends the period of the pseudo-random numbers and increases the randomness of the output signal. Suppose multiple $m$ sequences are used with identical sequences. In this case, the pseudo-random numbers generated by each $m$ sequence will have the same pattern of change and the randomness of the signal will be reduced. Therefore, multiple LFSRs with different sequences will generate multiple $m$ sequences simultaneously. Since the parameter lookup table in this device has a storage depth of 4k and a width of 12, 12 LFSRs with different steps are used to generate 12 $m$ sequences simultaneously. For each $m$ sequence, the most significant bit is taken and the 12 data are taken as a 12-bit pseudo-random number. The pseudo-random numbers generated in this way have a longer period and can be better adapted to requirements. Assuming that the series of the 12 $m$ sequences are $n_{1}$, $n_{2}$, ...$n_{12}$, the period $T$ of the 12-bit pseudo-random numbers they generate is given in Eq.(\ref{eq:4}).

\begin{equation}
\begin{aligned}
T & = \left ( 2^{n_{1} }  -1\right ) \left ( 2^{n_{2} }  -1\right ) \cdots \left ( 2^{n_{12} }  -1\right )\\
  & = \prod_{i=1}^{12} \left ( 2^{n_{i} }  -1\right )
\label{eq:4}
\end{aligned}
\end{equation}

\subsection{Data encoding}
\begin{figure}[h]
\centering
\includegraphics[ width=7cm]{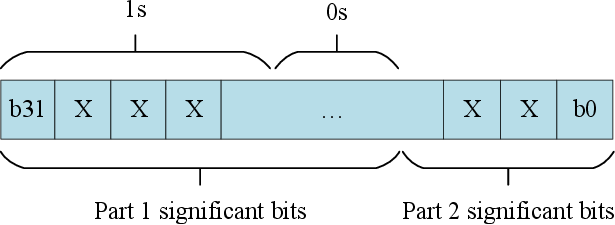}
\caption{Data frame structure}
\label{fig:frame}
\end{figure}

\begin{figure}[h]
\centering
\includegraphics[ width=8cm]{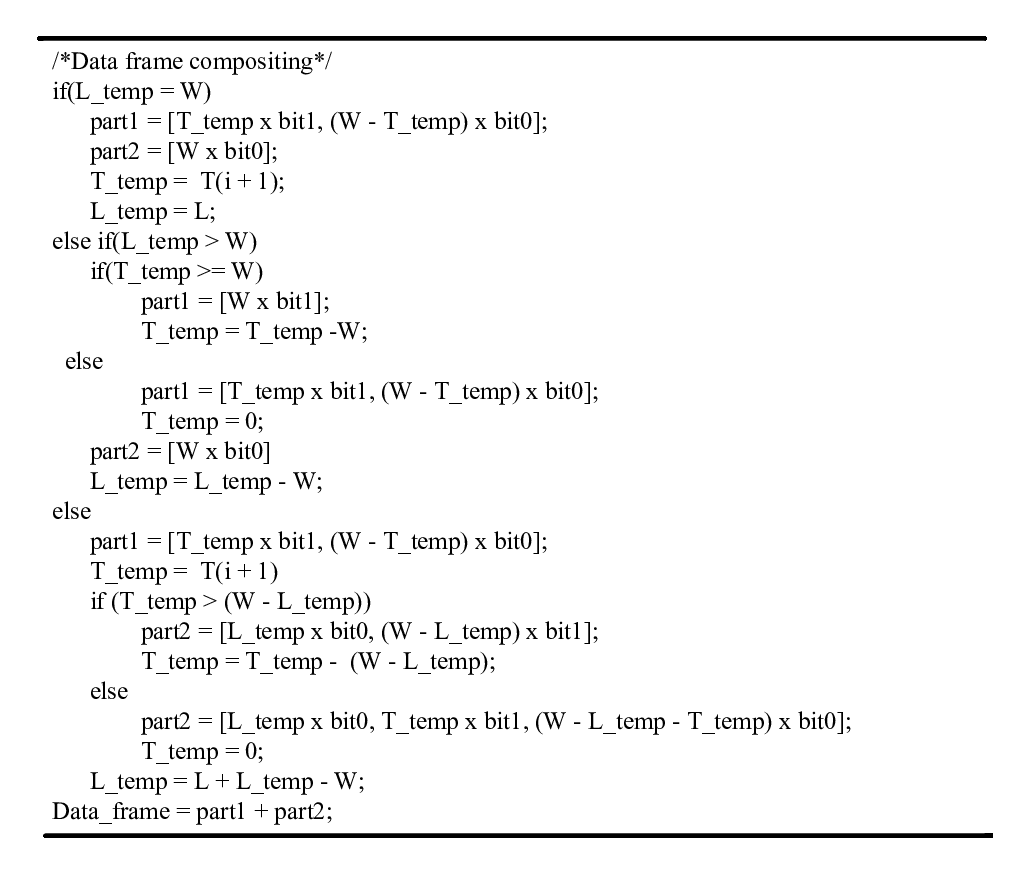}
\caption{Pseudocode of timing parameter decoding.}
\label{fig:encode}
\end{figure}

Each time parameter represents the time interval of the DTC output, which is the number of 1s in the high-speed serial data. The data encoding module encodes the time parameters into several parallel data. Each parallel data is called a data frame, and the composition of 1's and 0's in each frame is determined by the encoding module calculation. The bit width of a data frame is denoted as $W$ (32-bit in this design), the total duration length of each timing signal is denoted as $L$, and the i-th time parameter is denoted as $T_{i}$. In order to achieve continuous time parameter encoding, the data encoding module composes the data frames into two parts, Part1 and Part2. As shown in Fig.\ref{fig:frame}, Both have a bit width of 32-bit, but each part's significant bits range varies with the encoding. Part1 represents the data composited in the current timing signal, the high part of which is valid, and the remaining low part is invalid. Part2 represents the data composited in the next timing signal, the low part of which is valid and the high part is invalid. The sum of the valid bit widths of the two parts is $W$. As shown in Fig.\ref{fig:encode}, the synthesis process of the data frame is divided into three different cases, in which the number of bits remaining in the timing signal is recorded as $L\_temp$; the number of bits remaining in the time interval is recorded as $T\_temp$.

\begin{figure}[h]
\centering
\includegraphics[ width=6cm]{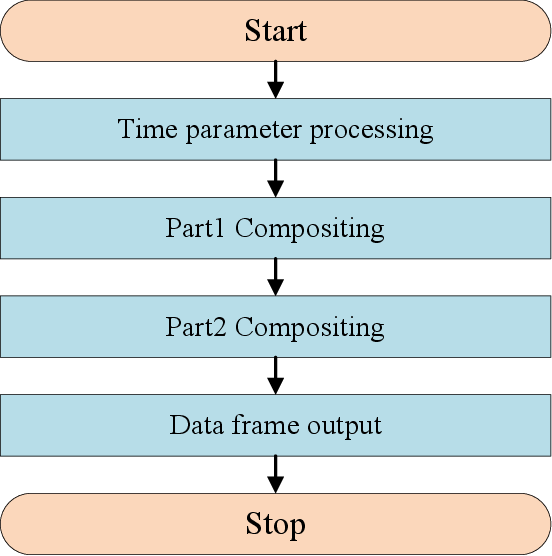}
\caption{Flowchart of data frame compositing.}
\label{fig:flow}
\end{figure}

If the frame just completes i-th timing signal encoding. At this time, all 32 bits of Part1 are valid, and according to the value of $T\_temp$, Part1 is set to 1; Part2 does not need to be set to 1. If the frame cannot complete i-th timing signal encoding, $L\_temp$ is greater than $W$. At this time, all bits of Part1 are valid; according to the value of $T\_temp$, Part1 is set to 1; there is no need to splice, and Part2 is set to 0. If this frame can complete i-th timing signal encoding, spare bits are needed for the next time parameter to participate in the encoding, that is, $L\_temp$ is less than $W$. According to the value of $L\_temp$, divide the valid bits of Part1 and Part2, and then set Part1's valid bits to 1 according to the value of $T\_temp$, which needs to be spliced. So update $T\_temp$ to T(i+1), and set the valid bits of Part2 to 1 according to the value of $T\_temp$. Finally, Part1 and Part2 are added to complete encoding one data frame.

\subsection{Serdes}

SerDes module converts the parallel data into high-speed serial data and generates timing signals. The SerDes module comprises a Physical Media Attachment (PMA) and a Physical Coding Sublayer (PCS). The PMA and PCS sublayers internally integrate high-speed serial-to-parallel conversion circuits, clock generation circuits, 8B/10B coder and decoder circuits, and channel binding circuits. These circuit structures provide the physical layer foundation for high-speed serial data transfer protocols.

\begin{figure}[h]
\centering
    \includegraphics[ width=8cm]{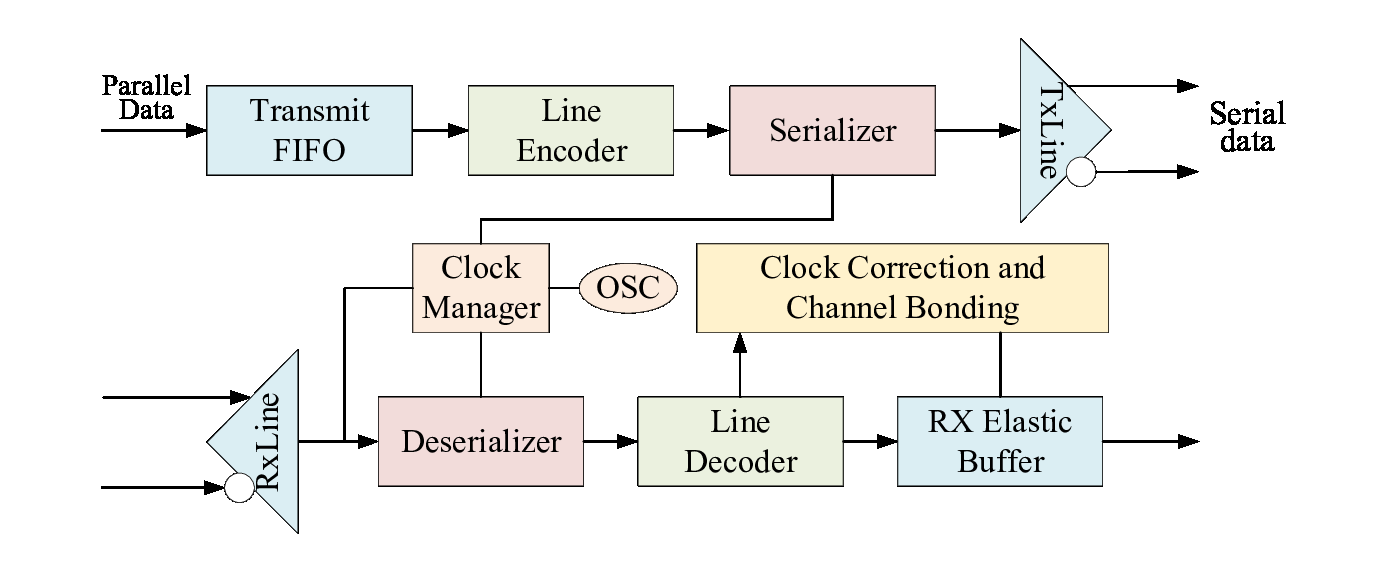}
    \caption{Structure of SerDes Module.}
    \label{fig:serdes}
\end{figure}

The structure of the SerDes module in the FPGA is shown in Fig.\ref{fig:serdes}. The SerDes module structure includes two channels: the transmit line (TX Line) and the receive line (RX Line). The main thing used in this DTC device is the TX line. The parallel data generated by the data encoding module is fed from the Transmit FIFO to the transmitter module in the SerDes fabric, and the data passes through the encoder line for 8B/10B encoding. To keep the data original, the 8B/10B encoding function will not be used in this design. Then, the data is entered into the transmit buffer. The primary clock isolation in the transmit buffer is between the two clock domains of the PMA sublayer and the PCS sublayer, which is used to solve the problem of clock rate matching and phase difference between the two. The Serializer converts the final data to high-speed serial data and sends it out.	

\section{Experiment and result}

\begin{figure}[h]
\centering
\includegraphics[ width=8cm]{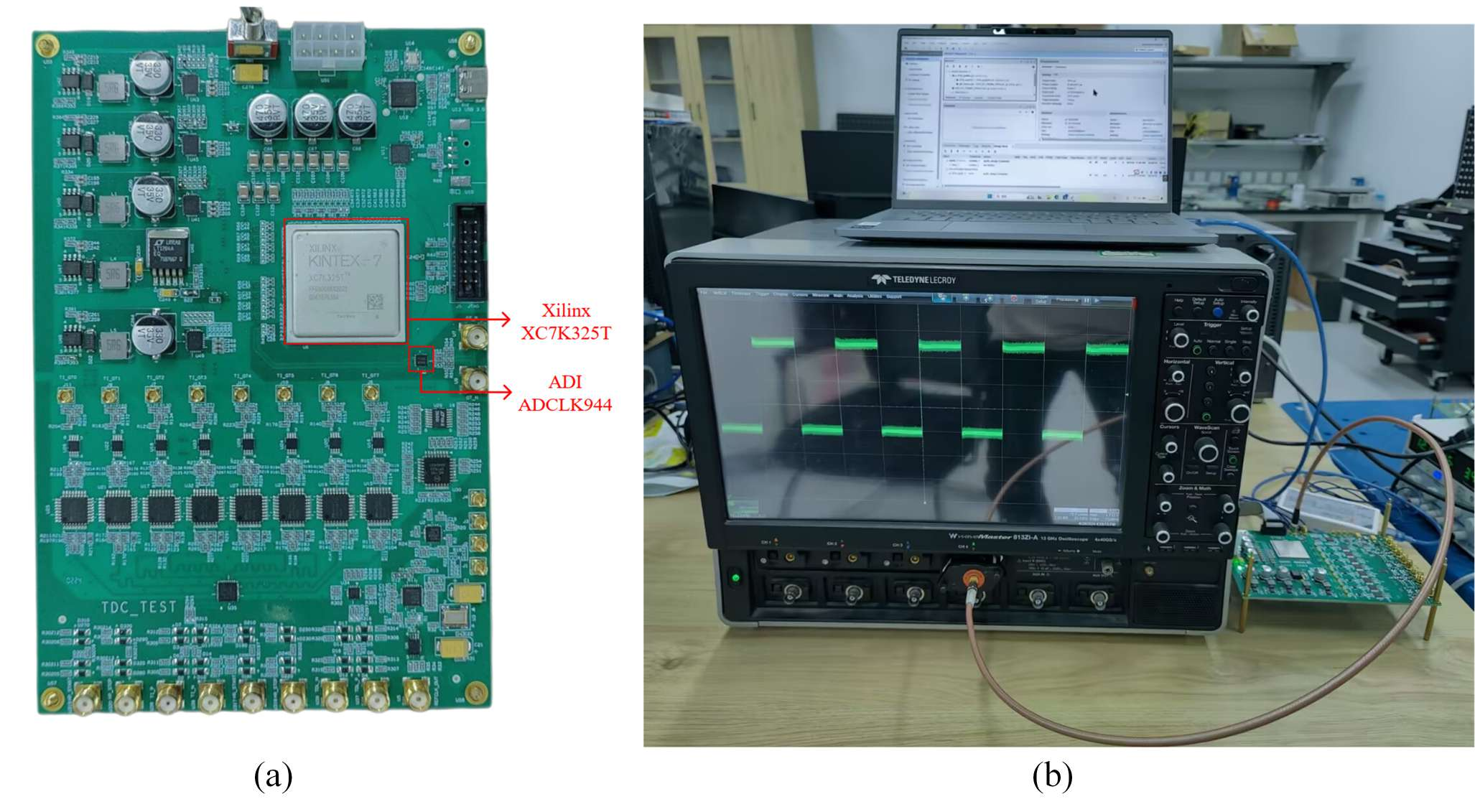}
\caption{Photograph of the experimental platform. (a) PCB. (b) Experimental platform.}
 \label{fig:PCB}
\end{figure}

In order to verify the feasibility and performance of this set of designs, we set up an experimental platform based on Xilinx FPGA chips. Fig.\ref{fig:PCB} illustrates that the core is formed by the Xilinx FPGA chip Kintex-7 specific model xc7k325t and ADI's clock fanout, ADCLK944. The clock fanout operates at a maximum frequency of 7.0 GHz with a wideband random jitter of 50 fs. The output timing signals are observed using a Teledyne LeCroy 813Zi-A oscilloscope with a bandwidth of 13 GHz and a maximum sample rate of 40 GSPS. In this design, the serial output rate of the high-speed transceiver interface is set to 10 Gbps by adjusting the clock source, so the resolution is 100 ps.

Table \ref{Resource} shows the FPGA resources consumed by a program written in verilog, synthesized and implemented. Lookup table (LUT), FF (Flip-Flop), and slices are the basic resources of the FPGA chip. Gigabyte Transceive (GTX) is the high-speed transceiver interface (SerDes) of this FPGA chip.

\begin{table}
\centering
\caption{Resource consumption}
\label{Resource}
\resizebox{0.35\textwidth}{!}{\begin{tabular}{lccr}
\toprule 
Resource & Generation & Encoding & SerDes\\
\midrule 
LUT      & 38 & 511 & 89 \\
FF & 952 & 4176 & 1064\\
Slice    & 28 & 193 & 48\\
GTX      & 0 & 0 & 1\\
\bottomrule
\end{tabular}
}
\end{table}

\begin{table*}[h]
\centering
\caption{Comparison of timing signal generator}
\label{DTC}
\resizebox{\textwidth}{!}{%
\begin{tabular}{c c c c c c c c}
\toprule
Ref,year & Method & Resolution (ps) & INL (LSB) & DNL (LSB) & Dynamic Range & Precision (ps) & Multi-function\\
\midrule 
This work  & SerDes in FPGA & 100 & -0.04/0.03 & -0.02/0.02 & 40 $\mu s$ & <3 & Random/Sequences\\
Ref\cite{delay2},2021 & Dsp blocks-based delay line & 14.20 & -19.63/18.95 & -1.00/22.08 & 10.86 ns & <2.80 & N/A\\
Ref\cite{compare1},2021 & Multiple delay lines & 11.30 & -1.00/1.94 & -7.01/2.05 & 0.1 s & N/A & N/A\\
Ref\cite{count3},2021 & Delay line & 16 & N/A & N/A & 1 s & <470 & N/A\\
Ref\cite{compare2},2021 & Direct Digital Synthesis & <10 & N/A & N/A & >7.5 ns & $\le$60 & PWM\\
Ref\cite{compare3},2019 & Multi-stage time interpolation & 3 & -3.0/1.23 & -0.7/0.63 & 5 s & 3.2 & Amplitude adjust\\
Ref\cite{delay3},2018 & Time folding and interpolating & 5 & -0.14/0.18 & -0.18/0.2 & 10 s & 9.31 & N/A\\
Ref\cite{delay1},2018 & Delay line & 19.40 & -1/6 & -1/3.95 & 4.95 ns & <4.5 & N/A\\
\bottomrule
\end{tabular}
}

\end{table*}

\begin{figure}
\centering
\includegraphics[ width=8cm]{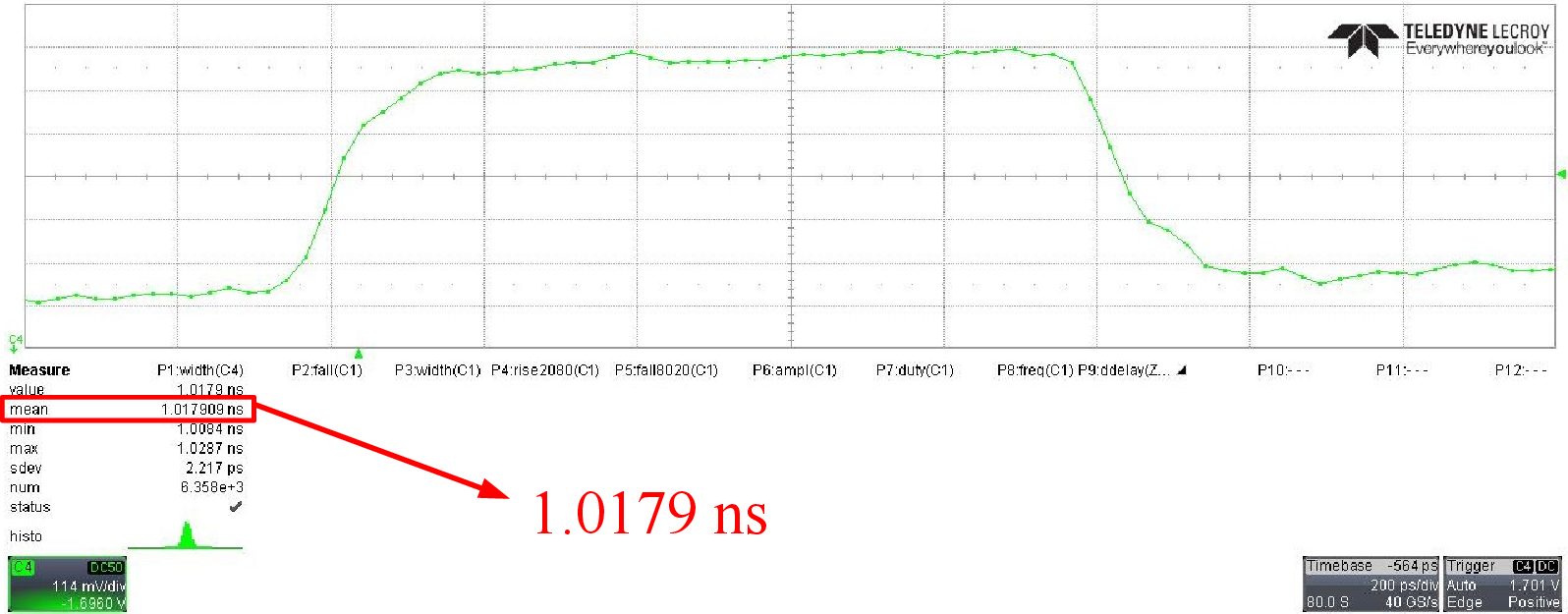}
\caption{Minimum timing interval test chart.}
 \label{fig:mini}
\end{figure}
\begin{figure}
\centering
\includegraphics[ width=8cm]{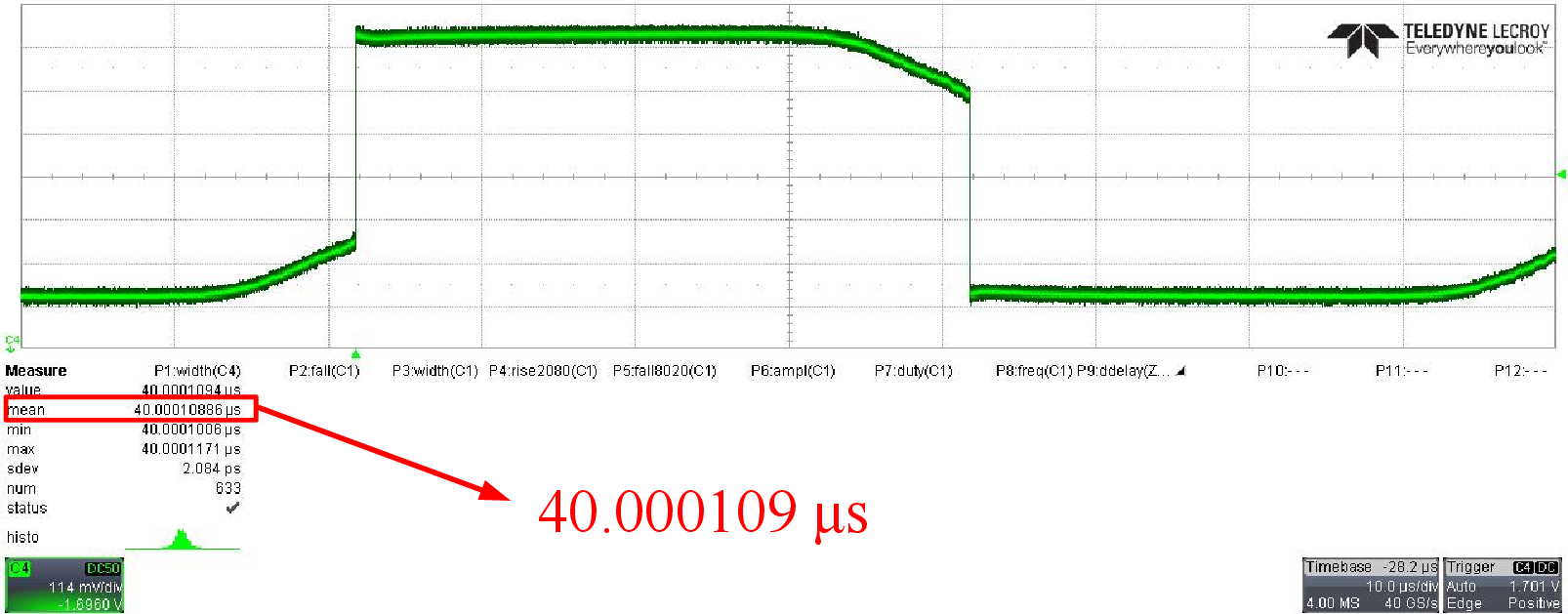}
\caption{Maxmum timing interval test chart.}
 \label{fig:max}
\end{figure}

\begin{figure}
\centering
\includegraphics[ width=8cm]{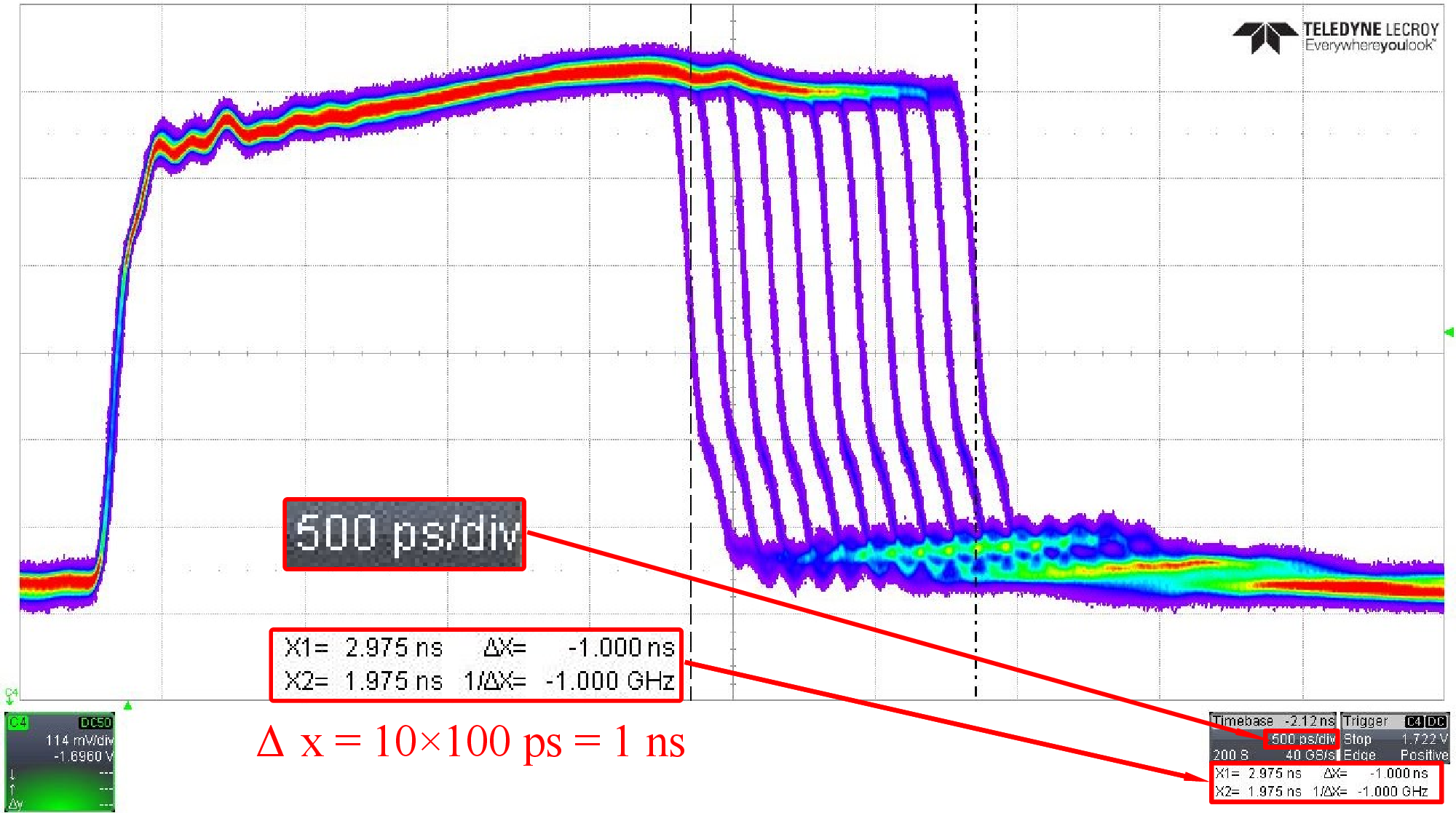}
\caption{Timing resolution test chart.}
 \label{fig:res}
\end{figure}

Fig.~\ref{fig:mini} shows the minimum time interval, with a minimum interval of 1.0179 ns and the standard deviation of the signal is 2.22 ps. The maximum time interval is shown in Fig.~\ref{fig:max}, with a maximum interval of 40.000109 $\mu$$s$, and the standard deviation of the signal is 2.08 ps. As the signal synthesis precision is tied to the clock performance, the precision of the time signal has good consistency over the entire dynamic range, with a standard deviation of less than 3 ps. Fig.\ref{fig:res} shows the timing signal resolution reaching 100 ps.

\begin{figure}
    \centering
    \subfigure[DNL test chart.]{
        \includegraphics[width=0.22\textwidth]{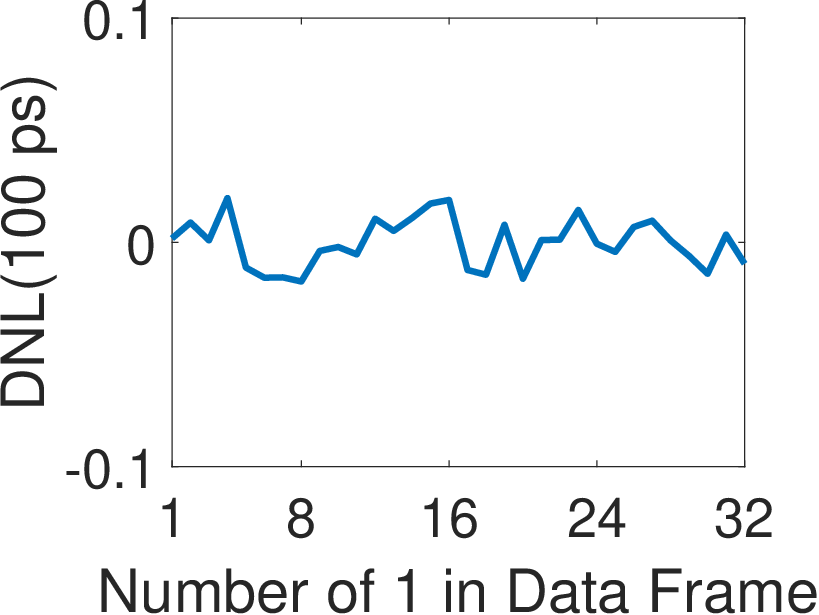}
        \label{fig:DNL}
    }
    \subfigure[INL test chart.]{
        \includegraphics[width=0.22\textwidth]{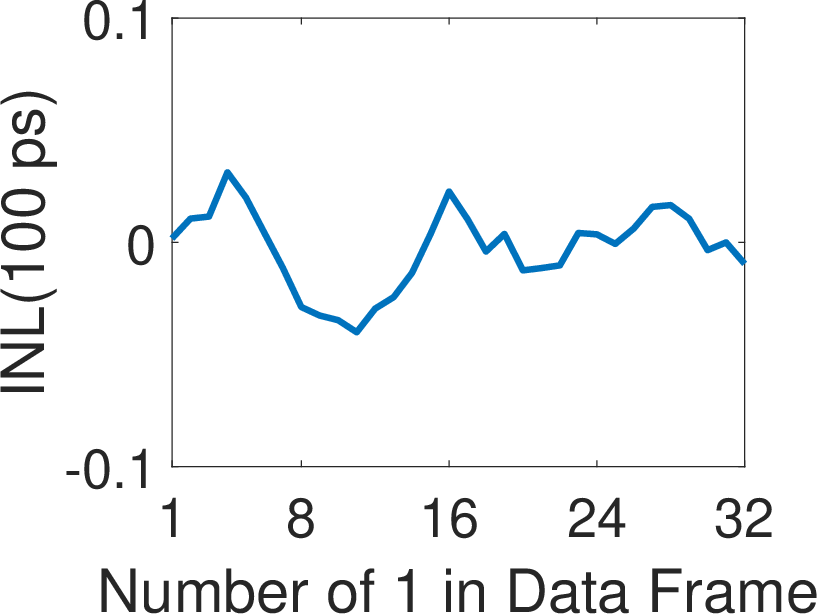}
        \label{fig:INL}
    }
    \caption{Nonlinearity test chart of data frame.}
   
\end{figure}

The synthesis of the 32-bit data frame is at the heart of timing signal synthesis, and its nonlinear performance is representative of DTC performance over the entire dynamic range. We tested the INL and DNL of the data frame. Fig.~\ref{fig:DNL} shows the DNL, which ranges from -0.02 to 0.02 LSB. Fig.~\ref{fig:INL} shows the INL, which ranges from -0.04 to 0.03 LSB.

\begin{figure}
\centering
\includegraphics[ width=8cm]{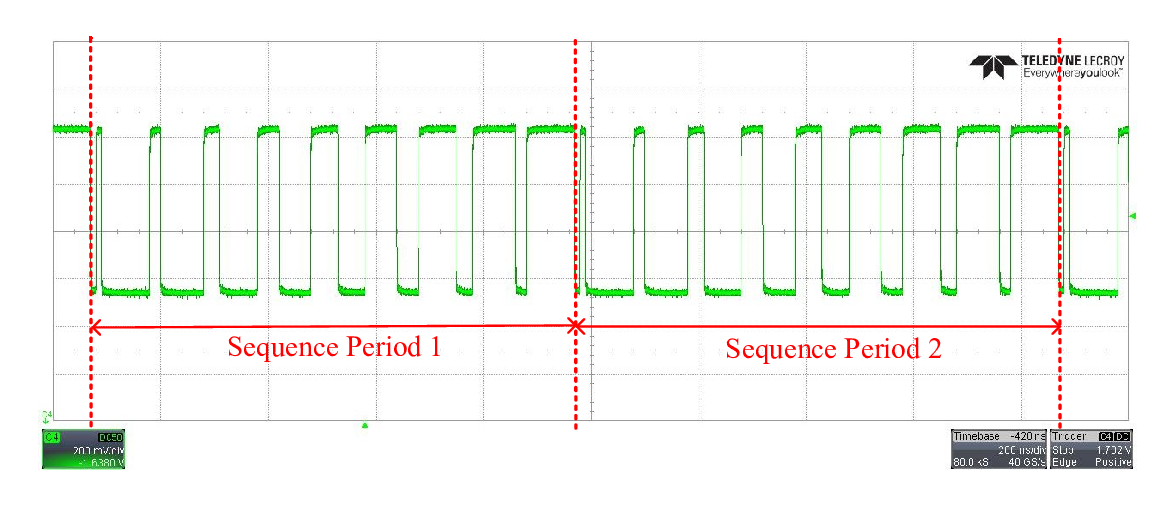}
\caption{Timing sequences test chart.}
 \label{fig:sequence}
\end{figure}
\begin{figure}[t]
\centering
\includegraphics[ width=8cm]{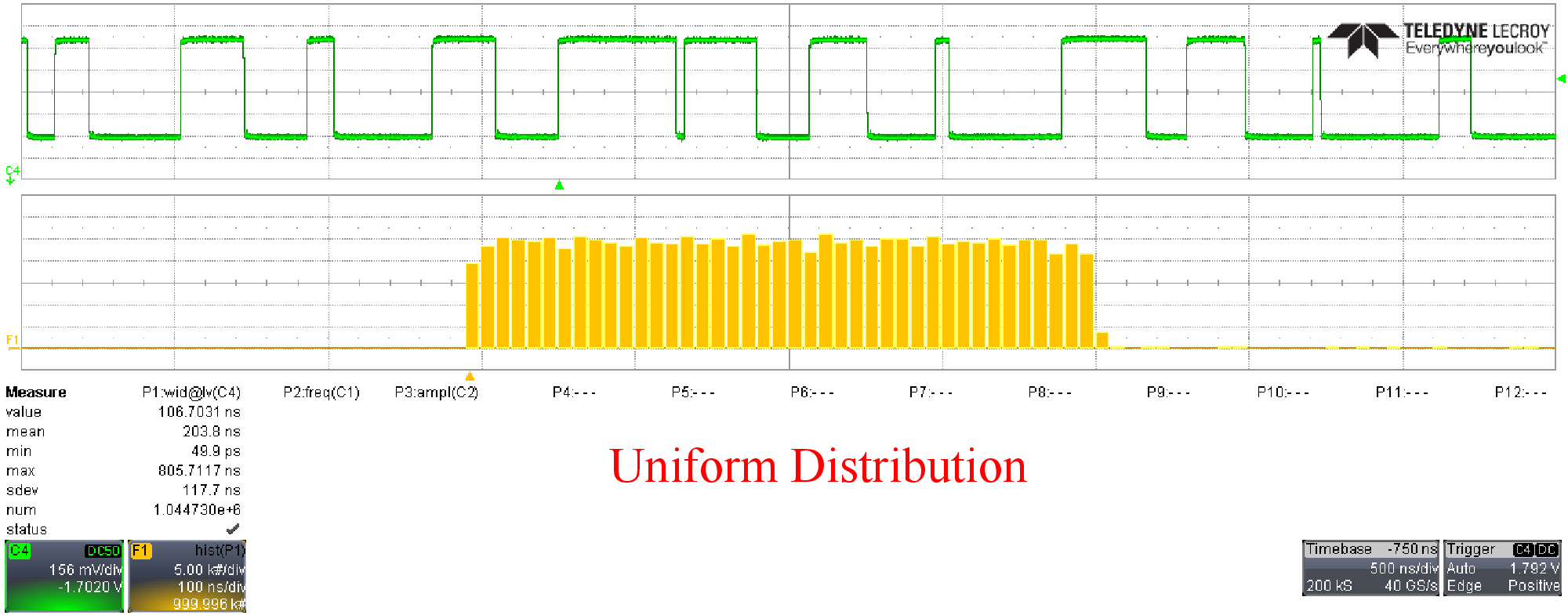}
\caption{Random timing signal test chart.}
 \label{fig:random}
\end{figure}

Fig.~\ref{fig:sequence} shows the sequence generation function, demonstrating two sequence cycles. The sequence increases in pulse width with time, with nine timing signal as the cycle. Users can adjust the number of timing signal and interval according to their needs. After sampling 1 M random intervals, the random timing signal function is shown in Fig.~\ref{fig:random}. According to the statistical results, it can be concluded that the random timing signal obeys a uniform distribution.

Table~\ref{DTC} shows a comparison of timing signal generator designs published in recent years with this study. The comparison of several key performance metrics, such as resolution and nonlinear performance. This method achieves excellent nonlinear performance compared to other methods. It has a larger dynamic range than \cite{delay2,compare2,delay1}. In terms of resolution, there is no advantage but high relative precision is achieved without calibration. In addition, this DTC has  multifunctional output ability, which is not available in other works. Last, the resolution and dynamic range of this DTC can be flexibly reconfigured and can reach higher upper limits as the FPGA performance increases.

\section{Conclusion}

This paper presents a novel DTC leveraging FPGA high-speed transceivers for high-precision timing signal synthesis. The design achieves an INL of [-0.04, 0.03] LSB and a DNL of [-0.02, 0.02] LSB. A real-time encoding scheme converts time information into code data, reducing FPGA memory usage and improving dynamic range. The DTC supports both timing sequence and random signal outputs, making it versatile for time-domain measurement systems. FPGA high-speed transceivers enable precise timing signal generation without calibration, with support for variable resolution. Real-time encoding eliminates the need for large memory storage, conserving FPGA resources. It supports timing sequences and random signals and is adaptable to applications like LiDAR, PET-CT, and other time-domain test systems.

\end{document}